\documentclass[conference]{IEEEtran}
\ifCLASSINFOpdf
\else
\fi
\usepackage{geometry}
\usepackage{epsfig}
\usepackage{array}
\usepackage{amssymb}
\usepackage{amsfonts}
\usepackage{verbatim}
\usepackage{url}
\usepackage{amsmath}
\usepackage{xcolor}
\usepackage[justification=centering]{caption}
\usepackage{subcaption}
\usepackage{wrapfig}
\usepackage[numbers,sort&compress]{natbib}
\usepackage{enumitem}
\usepackage{algorithm}
\usepackage{algpseudocode}

\usepackage{ifthen}
\usepackage{calc}
\usepackage{pifont}
\usepackage{color}
\usepackage{fancyhdr}
\usepackage{xspace}
\usepackage{xkeyval}
\usepackage{multirow}

\newcounter{hours}
\newcounter{minutes}

\newboolean{timeofmake}
\setboolean{timeofmake}{false}

\newcommand{\ignore}[1]{}

\newcommand{\dontinclude}[1]{ }

\newcommand{\name}{{Slytherin}\xspace}

\begin{document}
\title{\name: Dynamic, Network-assisted Prioritization
of Tail Packets in Datacenter Networks}

\author{\IEEEauthorblockN{Hamed Rezaei}
\IEEEauthorblockA{University of Illinois at Chicago, USA\\
Email: hrezae2@uic.edu}
\and
\IEEEauthorblockN{Mojtaba Malekpourshahraki}
\IEEEauthorblockA{University of Illinois at Chicago, USA\\
Email: mmalek3@uic.edu}
\and
\IEEEauthorblockN{Balajee Vamanan}
\IEEEauthorblockA{University of Illinois at Chicago, USA\\
Email: bvamanan@uic.edu}}


%Telephone: (800) 555--1212, Fax: (888) 555--1212}


\maketitle

\begin{abstract}
Datacenter applications demand both low latency and high throughput; 
while interactive applications (e.g., Web Search) demand 
low tail latency for their short messages 
due to their partition-aggregate software architecture, 
many data-intensive applications (e.g., Map-Reduce) require 
high throughput for long flows as 
they move vast amounts of data across the network.
Recent proposals improve latency of short flows and 
throughput of long flows by addressing the shortcomings 
of existing packet scheduling and congestion control algorithms, 
respectively.
We make the key observation that long tails in the Flow Completion 
Times (FCT) of short flows result from packets that suffer congestion at 
more than one switch along their paths in the network. 
Our proposal, \textit{\name}, specifically targets packets that 
suffered from congestion at multiple points and 
prioritizes them in the network.
\name leverages ECN mechanism which is widely used in 
existing datacenters to identify such tail packets and 
dynamically prioritizes them using existing priority queues. 
As compared to existing state-of-the-art packet scheduling proposals, 
\name achieves 18.6\% lower $99^{th}$ percentile 
flow completion times for short flows 
without any loss of throughput. Further, \name drastically
reduces $99^{th}$ percentile queue length in switches 
by a factor of about 2x on average.
\end{abstract}


% \fi
\IEEEpeerreviewmaketitle

\section{Introduction}
\label{sec:intro}

Datacenters have emerged as the de facto platform for hosting user facing applications
that query vast amounts of Internet data (e.g., Web search) \cite{barroso1, barroso2}.
To provide efficient and up-to-date access to data for 
these foreground applications,
datacenters also run other background applications (e.g., Web crawler), which 
reorganize and update data.
The nature of these two broad categories of applications determine the 
traffic dynamics and objectives of the underlying datacenter network.

Foreground applications such as Web search require access to data 
that is spread across a large number of servers, for each user query. 
Each query must wait for responses from 
most of the servers (e.g., 99\% of servers)
to achieve a good tradeoff between query response time and quality 
(section \ref{sec:background}).
Thus, these applications generate relatively short flows (e.g., 8 - 32 KB) 
and are sensitive to the tail (e.g., $99^{th}$ percentile) flow completion times. 
On the other hand, background applications, by their very nature, 
generate long lasting flows which are sensitive to throughput. 
Therefore, a well-designed datacenter network must provide low 
tail flow completion times for short flows and high throughput for long flows. 

Load balancing, congestion control, and packet scheduling play a key role in 
determining the bottomline performance of datacenter networks.
Good load balancing is crucial for both latency and throughput; 
poor load balancing leads to congestion hotspots (i.e., long queues)   
that worsen (tail) latency and leads to under-utilization of 
network capacity (i.e., throughput loss). 
Fortunately, many recent proposals achieve near-perfect load balancing
\cite{presto16,CONGA,hula}. 
While congestion control proposals improve tail flow completion times 
to some extent, their main thrust lies in modulating the flow rate
over several RTTs using network feedback (e.g., RTT, ECN) 
without causing congestion \cite{dctcp,dcqcn,timely,d2tcp,d3}. 
As such, most congestion control approaches focus on 
long flows which last a few tens of RTTs. Because short flows only last for 
a handful of RTTs, packet scheduling plays a much more direct 
role in determining the flow completion times of short flows. 
Therefore, we focus on packet scheduling in this paper. 

Existing packet scheduling 
approaches~\cite{d3,d2tcp,karuna,pfabric,pias,pase,pdq} prioritize 
short flows, in an effort to mimic Shortest Job First (SJF) scheduling, 
which is known to minimize average flow completion times.
Information-aware flow scheduling approaches~\cite{d3,d2tcp,karuna,pfabric}
explicitly use flow sizes or deadline information to 
prioritize flow packets using multiple queues. 
Information-agnostic approaches~\cite{pias} gradually demote flows 
from higher priority queues (i.e., every flow would start at the highest priority 
queue and move down in priority after sending some packets). 
In this paper, we ask the question: 
\textit{Is it possible to further improve tail flow completion times 
beyond SJF scheduling?} 
We answer this question in the affirmative with \textit{\name}, 
a packet scheduling mechanism to improve tail flow completion times. 

It is well known that datacenter networks experience longer 
tail flow completion times (e.g., 5-10x of median) 
due to the bursty nature of traffic~\cite{benson}, 
shallow switch buffers~\cite{dctcp}, and 
partition-aggregate architecture of applications, which 
causes incast~\cite{d2tcp,timetrader}. 
We make the \textit{key} observation that a majority of 
packets that fall in the tail of the distribution 
incur long queuing delays at more than one switch in the 
network. \name's key idea is to prioritize those packets that have 
already incurred queuing earlier in their paths. However, realizing 
our idea requires tackling several implementation challenges.

Our \textit{first} challenge is to decide where to 
implement our scheme: in the end host or in the network. 
Because short flows do not offer enough time to detect and  
enforce priority at the end host, we opt for an in-network implementation. 
The \textit{second} challenge is to identify the right network signal from 
which we can reliably infer queuing delays. 
One naive way is for every switch to measure 
the waiting time in its queues and to include this waiting time 
in the packet header. However, this approach would require extra 
fields in the packet header and 
support for time stamping at the switches. 
Instead, we make a clever observation that Explicit Congestion Notification (ECN) marks
already provide this information at a 
somewhat coarse granularity. 
Because ECN is readily supported and widely deployed in today's datacenters~\cite{dctcp}, relying on ECN makes our design 
much more implementation friendly. 
Inspite of the ECN's coarser granularity, 
we found ECN marks to work well in practice. 
We observe that more than 50\% of tail packets 
incur ECN marks (incur large queuing)
at more than one switch at higher loads.  
(section~\ref{sec:Sneaker}, table~\ref{tab:opp}).
In \name, 
switches promotes packets that have ECN marks to higher priority queues. 
In contrast to existing schemes that 
implicitly or explicitly use static flow information (e.g., flow size, deadline) to infer priority, 
\name infers priority based on packet queuing 
delays, specifically targets 
packets that are more likely to fall in the tail, and 
improves tail flow completion times beyond SJF. 

In summary, \name's contributions include:
\begin{itemize}[noitemsep,topsep=0pt,parsep=0pt,partopsep=0pt]
\item Unlike prior schemes, \name specifically targets packets that are more likely to fall in the tail. 
\item \name infers queuing delays incurred by packets without costly 
timestamps but relies on clever use of ECN marks.  
\item \name \textit{drastically} reduces $99^{th}$ percentile 
queue length in switches by about a factor of 2x on average. 
\item \name achieves 18.6\% lower $99^{th}$ percentile 
flow completion times for short flows 
without any loss of throughput.

The rest of the paper is organized as follows. 
We begin with a background of datacenter applications and 
overview of packet scheduling approaches in section~\ref{sec:background}. We describe \name's design in 
section~\ref{sec:Sneaker}. Section~\ref{sec:methodology} and 
section~\ref{sec:results} present our methodology and results, 
respectively. Section~\ref{sec:relatedwork} presents an overview of 
other related work in this area and section~\ref{sec:conclusion}
concludes our paper.
\end{itemize}

 

\section{Background and motivation}
\label{sec:background}

%\begin{figure}
%\centering
%    \includegraphics[scale=0.50, trim=0cm 0cm 0cm 0cm, clip=true]{ICCCN18/figures/incast}
%\caption{\small{Partition-aggregate architecture}}
%\vspace{-0.3in}
%\label{fig:SDN}
%\end{figure}

To motivate our design, we first address two types of flow scheduling schemes in datacenters and then we show performance trade-offs with existing scheduling proposals. We also discuss these method's ability to improve $99^{th}$ percentile\footnote{we use tail FCT and $99^{th}$ percentile FCT interchangeably} flow completion time of flows which suffered from queuing delay at multiple points. 
\subsection{Information-agnostic scheduling:}
As we discussed in section \ref{sec:intro}, these methods try to schedule packets while there is no prior knowledge about the flow characteristics like flow size or flow deadline. In these methods, the scheduling algorithm usually assigns different priorities to packets and then each packet will be assigned to a specific queue based on the given priority. As an instance, one of past proposals, PIAS \cite{pias}, tries to leverage multiple priority queues to implement Multiple Level Feedback Queue (MLFQ), in which corresponding packets of a flow get gradually demoted from higher priority queues to lower ones based on bytes it has already sent \cite{pias}. 

The biggest advantage of information-agnostic scheduling schemes is ease of implementation which comes from not requiring prior hard-to-achieve information about the flows (e.g., flow size). Moreover, although they may add small complexities to switching fabric but since they don't rely on prior information about the flows, they can be implemented in real datacenter networks. 

The disadvantage of information-agnostic scheduling schemes is their inability to improve {\em tail} flow completion times. This is because, they schedule packets based on just limited information about the flow. More specifically, they try to treat each single packet (regardless of packet's history) based on a scheduling algorithm which can not discriminate among packets that suffered more queuing delay in previous hops (e.g., tail packets) and others. This is considered as a big problem because if packets of a flow experience different amount of queuing delay at multiple hops, flow completion times will be increased which is not tolerable in datacenter networks. 

As a conclusion, current information-agnostic scheduling methods cannot improve tail flow completion times because of their inability to discriminate among congested packets and normal packets. In general, information-agnostic schemes provide lower performance in contrast to information-aware methods. For example, $D^{3}$ \cite{d3} shows that as much as 7\% of flows may miss their deadlines with DCTCP \cite{dctcp} which is an information-agnostic scheme. However, as long as they don't require prior information about the flows, it's quite fair to not to expect them to provide as good performance as information-aware scheduling methods. We will discuss information-aware methods in the following section.

\subsection{Information-aware scheduling:}

Information-aware scheduling schemes try to provide higher performance by giving prior information about flows to switching fabric. In the other words, these schemes promise that switches know some key information about the flows like flow deadline, flow size or even flow remaining processing time. This information is usually given to switches either by a central controller or by end host applications. In general, we can divide these methods to two different groups:

\noindent
\subsubsection {Prior knowledge about flow deadlines}
These methods assume flow deadlines all are known apriori and switches greedily schedule flows with nearest deadline ahead of others. Having comprehensive knowledge about flow deadlines guarantees that almost non of deadlines are missed which drastically improves the performance of short, deadline sensitive flows. Some of methods which use this approach are $D^{3}$ \cite{d3}, $D^{2}TCP$ \cite{d2tcp}, and Karuna \cite{karuna}.

Remember from section \ref{sec:intro} that most of flows in datacenters are short and deadline sensitive. Furthermore, deadline aware schemes usually meet the requirements of vast fraction of flows in datacenters. However, although they have a big advantage over deadline-agnostic schemes but they still suffer from severe implementation issues. For many applications, such information (e.g., flow deadline) is difficult to obtain, and may even be unavailable \cite{pias}.     

\noindent
\subsubsection {Prior knowledge about flow sizes} These schemes assume flow sizes all are known to switches. They try to emulate Shortest Job First (SJF), which is known to minimize the average flow completion times. These methods greedily schedule shorter flows ahead of longer flows in their simplest form. In a more complicated form, they try to give higher priority to flows with shortest remaining size. Some of well known schemes that use this approach are pFabric \cite{pfabric}, PASE \cite{pase}, and PDQ \cite{pdq}.

Although information-aware scheduling methods provide better performance in contrast to information-agnostic approaches, but they try to use some information that are not easy to collect in datacenters. Prior proposals argue that flow sizes(or deadlines) can be tagged on packets by end host applications or a central controller may provide this information to switches. In both cases, we need to add extensive complexities to end-hosts and switches which is not only easy at all but even sometimes impossible.

As we saw, packet scheduling methods all suffer from either implementation issues or lack of smart scheduling decisions in case of scheduling congested tail packets. It turns out that most of current scheduling schemes are not effective enough or are very difficult to implement in real datacenters. Our biggest motivation is to provide a scheme which is both implementation friendly and smart enough to schedule mix flows while giving higher priorities to congested tail packets. To do so, we introduce {\em \textit{\name}} which is an information-agnostic method that not only improves tail flow completion times but even doesn't require any modifications at existing switches.

\section{\name}
\label{sec:Sneaker}
In this section, we discuss {\em \textit{\name}} which unlike prior schemes specifically targets packets that are more likely to fall in the tail. This is important because the performance of foreground datacenter applications (e.g., Web search) are sensitive to the tail of flow completion times. \name leverages Active Queue Management (AQM) schemes which are available in today commodity switches to identify congested tail packets and then assigns higher priorities to those packets. 

A TCP flow is not finished successfully unless all transmitted packets reach the destination. If some packets get delayed, the total flow completion time of flow will be increased. This problem caused by delayed tail packets of a flow because they face queuing delay at multiple switches on the path. Thus, although we may have most of packets arrived in a short time but delayed tail packets will increase the total flow completion time. 
Our main contribution is our \textit{novel} insight that 
packets that are more likely to fall the tail often incur 
congestion at multiple points in the network. 

To achieve our goal of identifying packets that incur congestion 
at more than one point in the network and prioritize them quickly 
to improve tail flow completion times, we need:\\
(1) A fast and a reasonably accurate signal to pinpoint packets 
that are likely to fall in the tail. 
We set out with the explicit goals of not requiring custom hardware and supporting coexistence with legacy transport protocols like TCP \cite{tcp}. To have higher response time, we should use in-network mechanisms to detect congestion at each switch independently.\\
(2) Provide a fast prioritization method to improve tail flow 
completion times. 
Consequently, the packet prioritization mechanism should be 
in-network because short flows only last for a few RTTs.  

\subsection{Identifying tail packets}
\begin{table}
\begin{center}
\caption{\name's opportunity \label{tab:opp}}
\begin{tabular}{ |c|c|c|c|c|c|c| } 
 \hline
 \textbf{Load}     & 40\%  & 50\%  & 60\% & 70\% & 80\% & 90\% \\ 
 \hline
 \textbf{Fraction of packets} & 1.3\% & 3.8\% & 19\% & 40\% & 56\% & 65\% \\
 \hline
\end{tabular}
\end{center}
%\caption{\name's opportunity \label{tab:opp}}
\end{table}

Explicit Congestion Notification (ECN) \cite{ecn} is widely used AQM scheme in datacenters. 
With ECN, switches mark packets when the queue length 
exceeds a certain predefined threshold. 
Instead of using expensive, unreliable packet timestamps, 
we leverage ECN to pinpoint tail packets. 
More specifically, we infer that marked packets have experienced congestion in their paths, and 
therefore, those packets must be prioritized ahead of 
packets without ECN marks. 

We performed an opportunity study to confirm our intuition that 
packets that are more likely to fall the tail often incur 
congestion at multiple points in the network and 
that we can identify those packets using ECN marks. 
For this study, 
we simulated typical datacenter traffic patterns, as reported 
in prior papers \cite{benson}, in a spine-leaf 
topology with 400 hosts (see section~\ref{sec:methodology}). 
All the hosts run DCTCP. 
Table~\ref{tab:opp} shows the fraction of 
tail packets 
(i.e., here we only consider packets whose flow completion times are greater than the $95^{th}$ 
percentile flow completion times; 
these packets are more likely to 
impact tail FCT than packets at lower 
percentiles)
that are ECN marked at 
more than one switch in their path. 
In other words, it shows 
what fraction of critical packets 
incur ECN marks at more than one switch. 
We clearly see that a \textit{significant} fraction of these 
packets are marked at multiple switches. 
This study clearly validates our approach. 

As long as each of switches perform ECN marking individually, congestion is recognized independently at each hop. This ensures that whether there is only one congestion point or multiple points of congestion, it will be reported by the corresponding switches to end hosts. {\name} uses this bit of information not to let the end hosts to decide about the congestion but react to it just in-network. More specifically, each {\name} switch check this bit of information individually to see whether the packet has experienced congestion at previous hops or not. 

Since AQM schemes are available in today commodity switches, we can simply leverage them to discriminate among packets that suffered from congestion in previous hops vs. other packets. 
Because ECN is meant to inform end hosts about congestion, 
existing switches do not check whether this bit is set or not. 
However, {\name} requires switches to check this bit and 
prioritize such packets.

\begin{figure}
\centering
\includegraphics[scale=0.30]{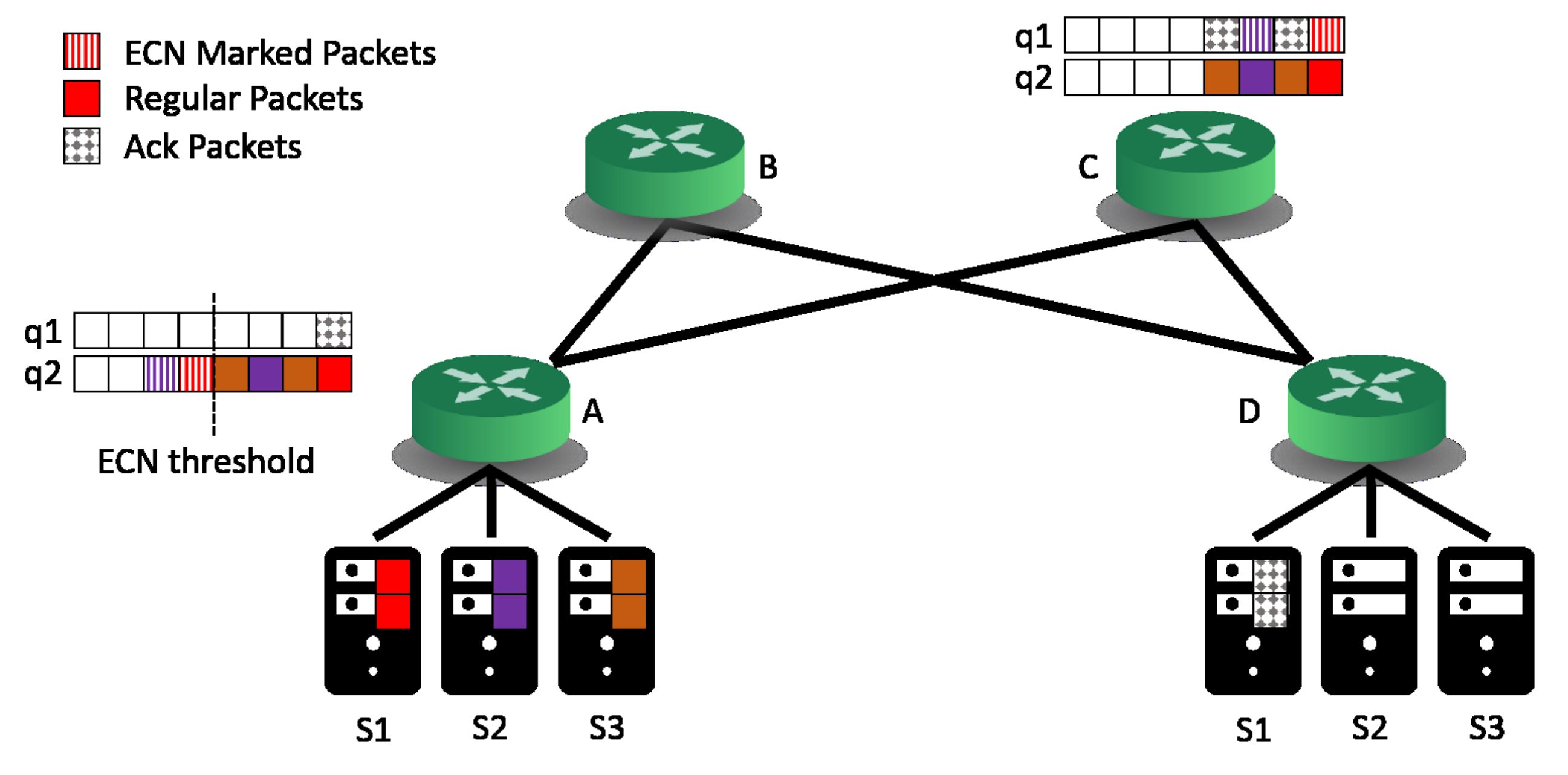}
\caption{\small{\name's high level idea}}\label{fig:ppe}
\label{fig:ex-nw}
\end{figure}

%\begin{figure}[!t]
%\centering
%\includegraphics[width=2.5in]{myfigure}
% where an .eps filename suffix will be assumed under latex, 
% and a .pdf suffix will be assumed for pdflatex; or what has been declared
% via \DeclareGraphicsExtensions.
%\caption{Simulation results for the network.}
%\label{fig_sim}
%\end{figure}

\subsection{Prioritizing tail packets}

As we discussed above, ECN is a widely used, in-network congestion notification mechanism at switches. Our analysis shows that this bit of information should be used by the switches to prioritize previously congested packets over the others. If the CE bit in the packet's header is set, this packet is considered as a congested packet; which means it requires priority escalation at next hop switches to minimize flow latency. Scheduling packets that suffered from congestion in the higher priority queue assures that tail packets that faced congestion earlier will not be delayed at the current switch. It significantly improves tail flow completion times which is a key performance metric in datacenters. 

{\name} switches require {\em two} queues per switch port. Packets are assigned to one of those queues based on their priority. The first queue is a priority queue which is dedicated to congestion experienced packets and the second one is a normal queue which is shared among all other packets regardless of their flow size or flow deadline. Both queues drain packets in a First-In-First-Out (FIFO) fashion and packets in the second queue get served if and only if the first queue is empty. Figure \ref{fig:ppe} shows \name's congestion detection and packet prioritization mechanism.

Alongside prioritizing congested packets, we prioritize ACK packets over other packets by scheduling them in the priority queue. Although it's not our novelty to prioritize ACK packets but our empirical analysis shows that ACK packets prioritization could be a good scheme to be used in conjunction with {\name}. 

\name is designed to prioritize packets \textit{after} the first hop
upon observing ECN marks.
{\name}'s Packet prioritization mechanism is totally application-agnostic. If a packet gets ECN marked, whether it's part of a short flow or a long flow, the packet will be prioritized at the next hop switch. Although prioritizing congested packets of short flows may be more efficient, 
it requires application knowledge and/or significant effort. 
Our experiments show that there is only a small degradation in performance in case of prioritizing both long and short flows in contrast to only prioritizing short flows.

\begin{algorithm}
 \caption{{\name}'s packet prioritization pseudocode}\label{alg:alg1}
\label{fig:ex-nw}
\begin{algorithmic}[1]
\For{Each packet "P" to be enqueued}
    \If{\textit{CE bit is set}}
       \State put "P" in higher priority queue
    \Else 
	    \State put "P" in lower priority queue
    \EndIf
\EndFor
\end{algorithmic}
\end{algorithm}

Algorithm \ref{alg:alg1} shows both congestion detection and packet prioritization steps at \name's switches.  {\name}'s design is such simple that its complete implementation requires about only 5 lines of code on the switches. Irrespective of where packets are ECN marked 
(i.e., enqueue vs. dequeue), 
Slytherin prioritizes marked packets to reduce 
flow completion times for both short and long flows. 
Overall, \name's implementation is simpler than 
other schemes such as Pfabric\cite{pfabric}, PIAS\cite{pias}, and PDQ\cite{pdq}.

\subsection{Parameters setting} \label{sec:form-other}
In this section, we will discuss important parameters settings that directly affect {\name}'s performance. 

\subsubsection{Transport protocol}
{\name} is designed to work in conjunction with DCTCP \cite{dctcp} as the underlying transport protocol. DCTCP elegantly aggregates the one-bit ECN feedback from multiple packets and multiple RTTs to form a multiple-bit, weighted-average metric for adjusting the window \cite{d2tcp}. Using this feedback, the senders adjust their congestion window sizes in a graceful manner, so that any congestion over the path from source to destination will be proactively treated.

As long as {\name} is an in-network scheme, it can work with any other transport protocols but we chose DCTCP due to the better throughput it provides for long flows and better flow completion time that it provides for short flows. Our analysis shows DCTCP's congestion window adjustment could be improved if ECN marked packets are drained faster at switches. In the other words, if ECN marked packets arrive faster, DCTCP adjusts its congestion window faster to avoid further congestion which leads to less packet drops, higher throughput, and lower flow completion times.

\subsubsection{ECN  threshold}
\name's most important parameter to tune is the ECN threshold at each switch queue. This threshold should be set carefully as it directly affects the performance. The threshold is set to 25\% of the total queue size. If the threshold is too short, many packets are assigned to the higher priority queue which means still many previously congested packets will be scheduled behind others in the current switch. Briefly, if the threshold is too short, {\name}'s performance will be degraded.

If the threshold is too big, probably no packet gets prioritized over others but only some packets that are about to drop. In this case, the switch marks a lot of packets as not-congested while they are potentially congested. It means that setting up a big threshold leads to performance degradation as well. We will study the sensitivity of our proposed method to ECN threshold in section \ref{sec:results}. 

\subsection{Fairness and high load scenarios}\label{sec:form-other}
It is important to discuss \name's fairness and its performance 
under sustained high load. 
Because \name only prioritizes a small fraction (i.e., tail packets) of flows, fairness is largely unaffected; our results show that Slytherin achieves better tail flow completion times for short flows and better throughput for long flows than PIAS and DCTCP. 
Another important question is
about \name's performance when there is simultaneously high queuing along most (all) switches along the path. 
Our experiments show that it is so rare for 
many switches along the path to have high queuing 
\textit{at the same time} 
that it does not affect
$99^{th}$ percentile FCT 
(e.g., such scenarios happen 
less than 1\% of the time). 

\begin{figure}
\centering
  \centering
  \includegraphics[width=\linewidth]{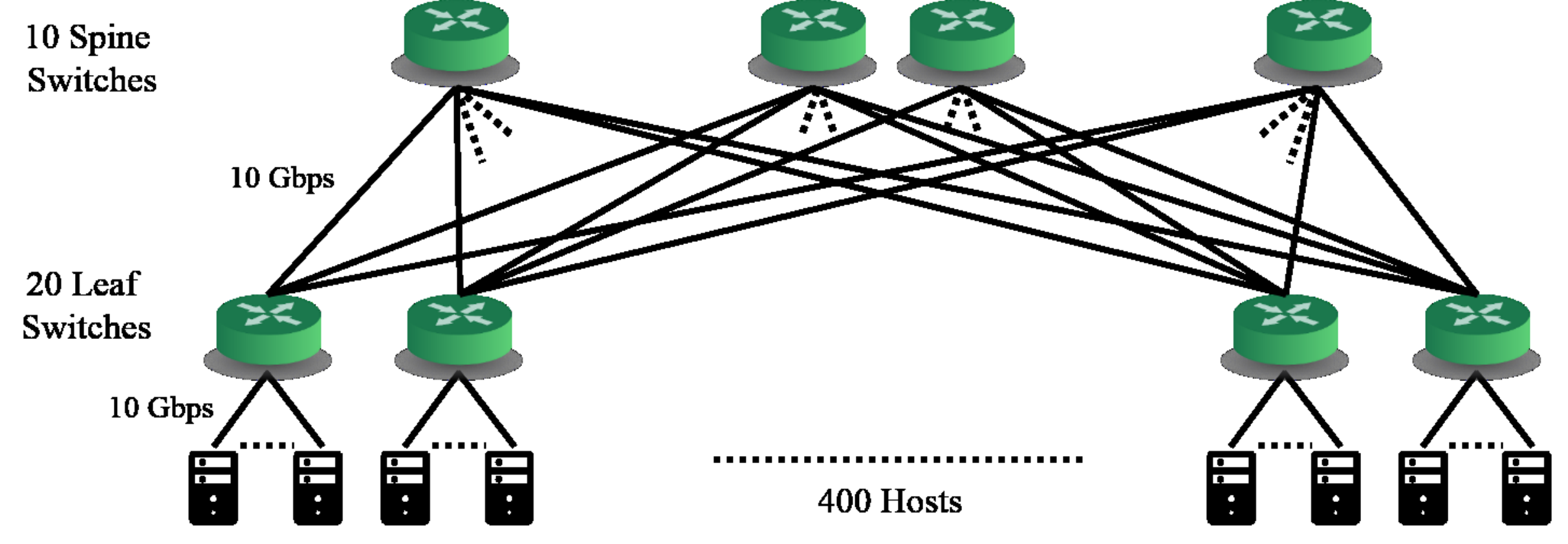}
  \label{fig:sub1}
\caption{Topology used in simulations}\label{fig:spineleaf}
\label{fig:test}
\end{figure}

\section{Experimental Methodology}
\label{sec:methodology}
In this section, 
we present the details of our 
simulator implementation, topology, and 
workload. 

\subsubsection{Topology}
We use ns-3 \cite{ns3} to simulate leaf-spine datacenter topology as shown in Figure \ref{fig:spineleaf}. Leaf-spine is a commonly used topology in modern datacenters \cite{pfabric}. In our simulations, the fabric interconnects 400 hosts through 20 leaf switches connected to 10 spine switches in a full mesh manner which provides over-subscription factor of 2. Each of leaf switches have 20 10 Gbps downlinks to the servers and 10 10 Gbps uplinks to the spine switches. The end-to-end Round-Trip Time (RTT) across the fabric is ∼80 $\mu$s. 

\subsubsection{Workload and Traffic} 
To evaluate our method, we simulate web search workload that is very common in modern datacenters. We consider two flow size distributions; short flows and long flows. All flows arrive according to Poisson process and the source and destination for each flow is chosen uniformly randomly. Short flow sizes are uniformly chosen in the range of 8 KB to 32 KB and our long flow's size is 1 MB. Since in web search workloads vast amount of all bytes are produced by 30\% of flows that are larger than 1 MB \cite{pfabric}, we use the same approach to produce the loads. We also use those short flows to produce incast type traffic which are quite common in web search workloads. 

To further evaluate \name's performance, we consider two metrics; one for short flows which is Flow Completion Time (FCT) and one for long flows which is throughput. Moreover, we check both \em{average} and \em{$99^{th}$} percentile flow completion times of short flows to measure the performance of our proposed method. Similarly, we evaluate \name's performance in both average and tail flow completion times in different incast scenarios. We use incast degrees of 24, 32, and 40 (number of parallel senders to a single receiver) to check the sensitivity of our method to incast degree. Next, we analyze \name's performance using other metrics such as queue length, convergence speed and number of reordered packets. 

\subsubsection{Compared schemes} 
\begin{itemize}

\item \textbf {DCTCP:}
We implemented DCTCP, 
capturing all details in their paper \cite{dctcp}. 
We use DCTCP as baseline. 

\item \textbf {PIAS:}  
We implemented PIAS on top of ns-3 \cite{ns3} simulator. 
The implementation assigns 4 queues to each switch port. 
At the very beginning, all flows get mapped to to the highest priority queue (queue 1). 
If number of sent bytes of a flow reach a threshold, the priority of corresponding flow will be decreased and then the rest of packets of the flow will be assigned to lower priority queues (queues 2 to 4). 
We set the ECN threshold to 25\% of the queue size and the load balancing method is flow ECMP.

\item \textbf {\name:} 
We implement {\name} on top of DCTCP \cite{dctcp}. 
Our implementation uses two queues per each switch port, 
recall from section \ref{sec:Sneaker} that 
we use one high priority queue for 
expediting ECN marked packets and ACKs, and one 
low priority queue for other packets. 
We set the ECN threshold to 25\% of the queue size and 
we use ECMP for load balancing. 
We set out the rest of our network 
configuration to match PIAS.

\item \textbf {Ideal SJF:}
We also implemented an ideal SJF scheduler.  
Our SJF scheduler is aware of flow sizes and  
maps short flows to higher priority queues. 
While this is not realistic as flow sizes are often not 
known apriori, we use our ideal SJF implementation
for deeper analysis of \name's queuing delays. 

\end{itemize}

As such, we use the same values for parameters 
that are common to DCTCP, PIAS, and 
\name (e.g., ECN threshold), and the values 
match those used in previous papers.
There are a number of packet scheduling approaches in the 
last few years such as pFabric \cite{pfabric}, PDQ \cite{pdq}, and PASE \cite{pase}. 
However, because PIAS compares to and outperforms these approaches, we only compare {\name} to PIAS.

\begin{figure*}
\begin{minipage}[b]{.64\linewidth}
\centering
\begin{subfigure}{.48\textwidth}
  \centering
  \includegraphics[width=\linewidth]{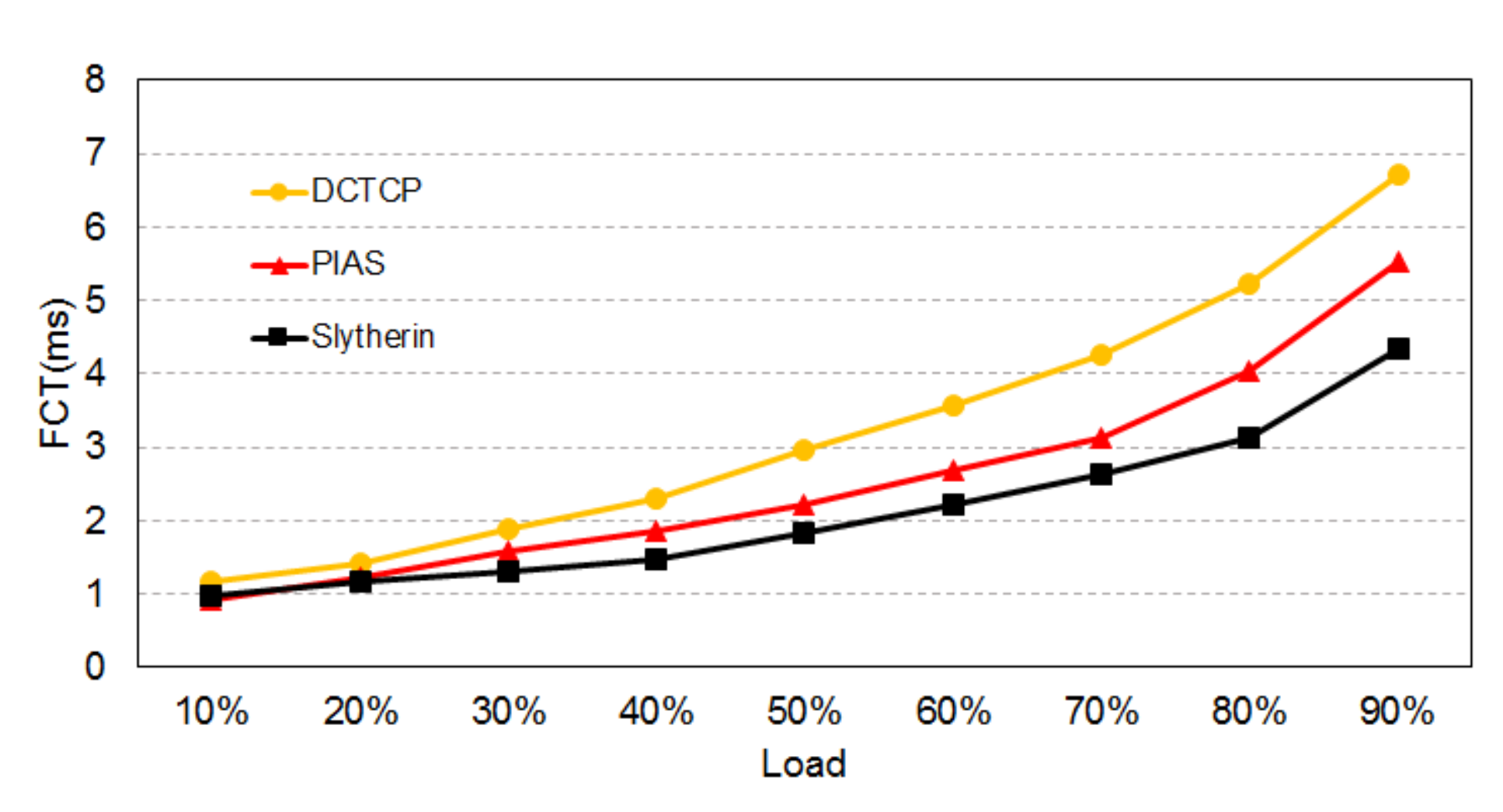}
  \caption{Tail ($99^{th}$) percentile FCT}\label{fig:FCT99}
\end{subfigure}%
\begin{subfigure}{.48\textwidth}
  \centering
  \includegraphics[width=\linewidth]{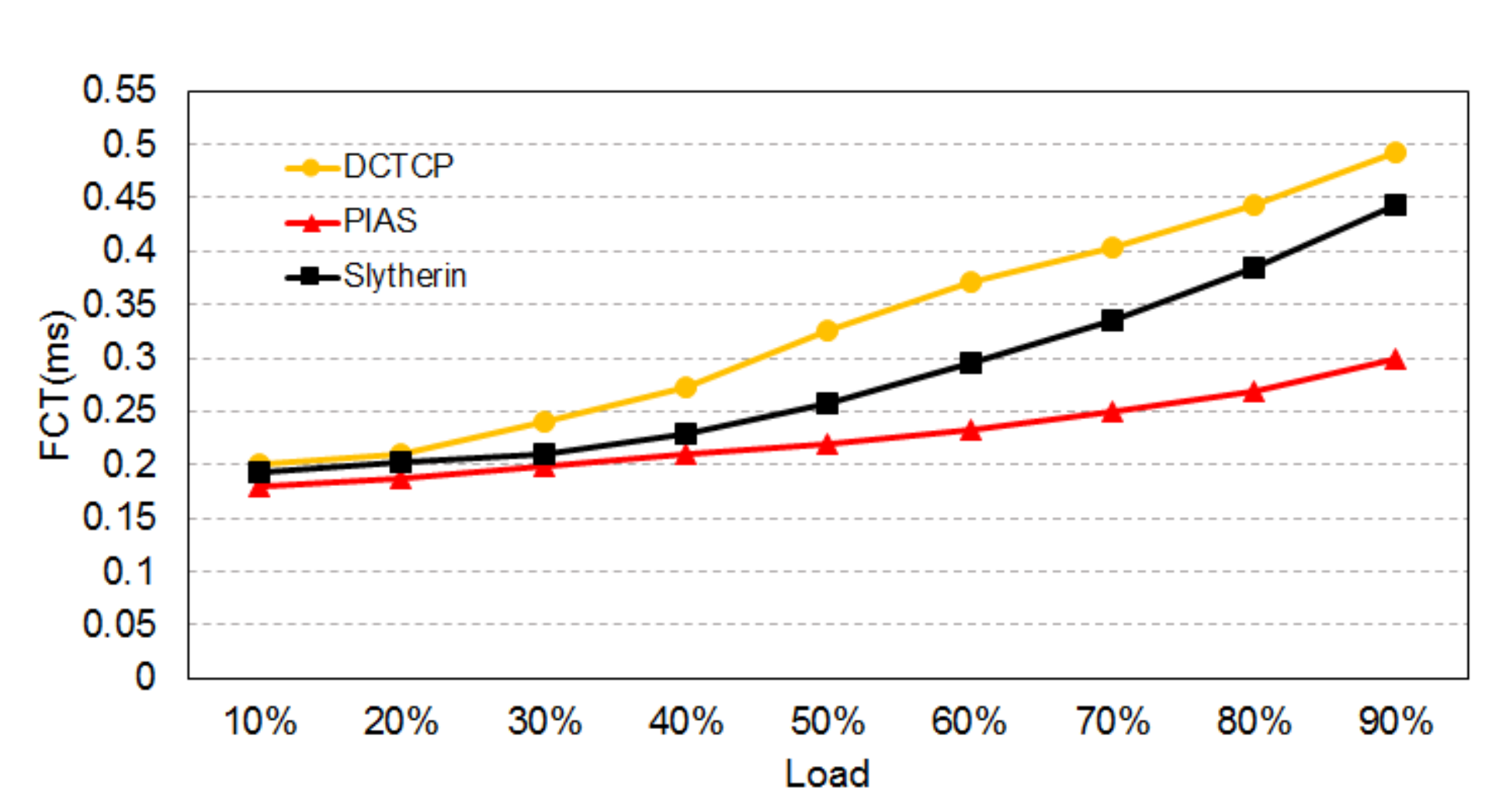}
  \caption{Average FCT}\label{fig:FCT50}
\end{subfigure}
\caption{Flow completion times (short flows)}\label{fig:comp}
\end{minipage}
\begin{minipage}[b]{.32\linewidth}
\centering
\includegraphics[scale=0.32,clip]{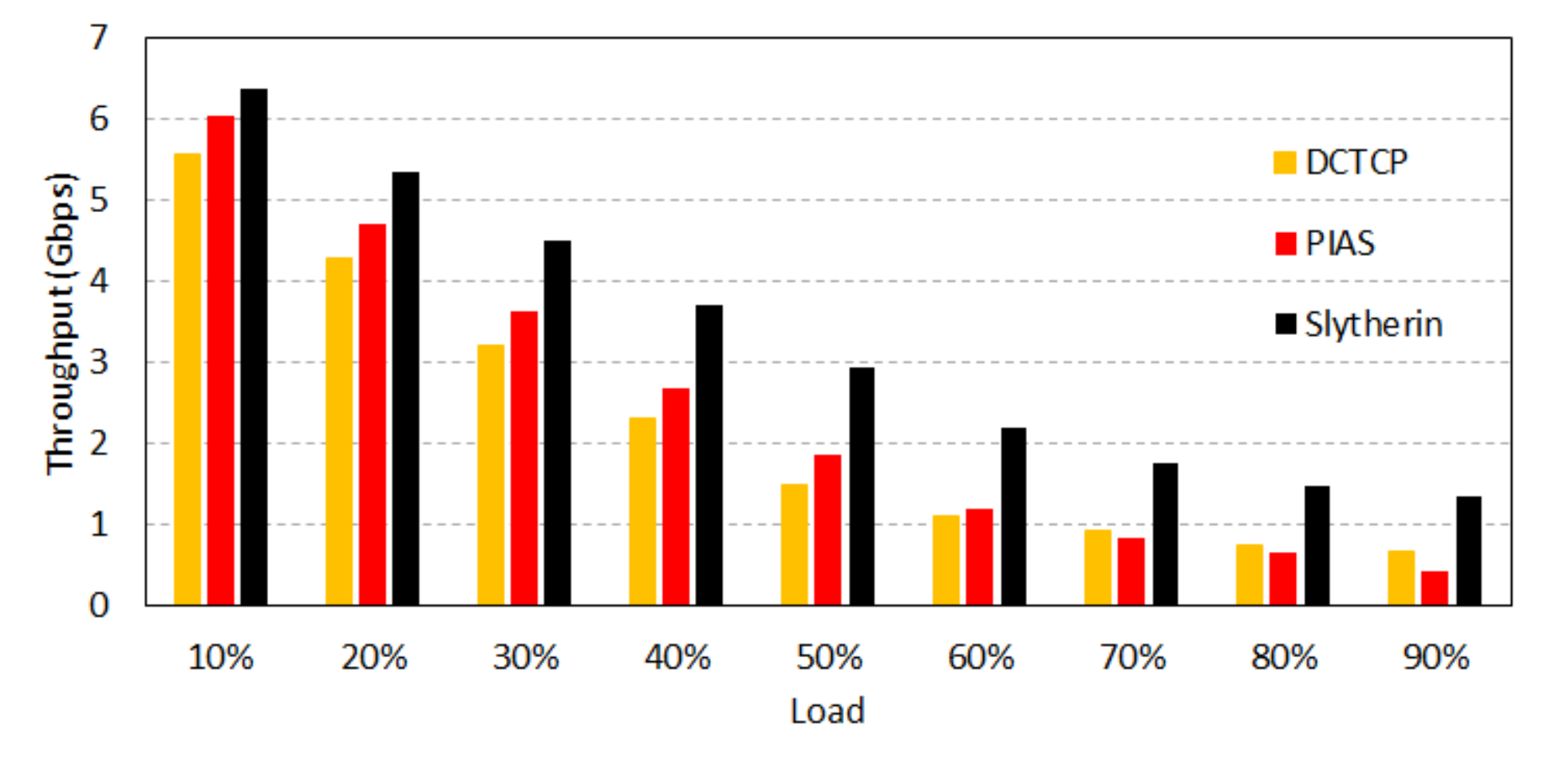}
\caption{Average throughput (long flows)}
\label{fig:Through99}
\end{minipage}%
\end{figure*}

\section{Results} 
\label{sec:results}
In this section, we evaluate the performance of {\name} in different scenarios on top of ns-3 \cite{ns3} simulator. 
Our performance evaluation consists of six parts: 
\begin{itemize}[noitemsep,topsep=0pt,parsep=0pt,partopsep=0pt]
    \item Bottomline comparison of \name's Flow Completion Time (FCT) and throughput to PIAS
    \item Analysis of \name's average queue length to explain our performance gains vs. PIAS and 
    ideal SJF. 
    \item Convergence analysis of DCTCP and \name
    \item Sensitivity to varying incast degrees 
    \item Sensitivity to ECN marking threshold 
    \item Analysis of packet reordering in {\name} vs. PIAS.
    
\end{itemize}

\subsection{Tail FCT and Throughput}
In this section we will show how {\name} performs in terms of both flow completion times (for short flows) and throughput (for long flows). The results for flow completion times and throughput are shown in figure \ref{fig:comp} and figure \ref{fig:Through99} respectively. In both figures, the X-axis shows the load factor on network. In figure 3 the Y-axis shows flow completion time in milliseconds and in figure 4 the Y-axis shows throughput in Gbps.

\subsubsection*{Flow Completion Time}
Figure \ref{fig:FCT99} compares the $99^{th}$ percentile completion times of {\name} and PIAS. 
As load increases, tail FCT increases for all the schemes.
PIAS and \name significantly outperform DCTCP. 
PIAS greedily assigns higher priority to flows that sent less packets regardless of queuing delays which leads to high queuing delay for some packets. 
Although PIAS benefits from ECN marking to prevent long queuing delay for longer flows, 
it falsely classifies some packets that have incurred higher queuing into 
lower priority queues which worsens tail FCT.
In contrast, \name targets tail packets and 
prioritizes them. 
At higher loads, \name achieves
better reduction in tail FCTs as there is 
more opportunity to schedule tail packets. 
\name achieves an average reduction in tail 
FCTs of about 20\% for loads greater than 20\% (typical 
operating point of most datacenters). 

Figure \ref{fig:FCT50} shows the {\em average} FCT for short flows. Although PIAS outperforms {\name} in average FCT but since the performance of foreground datacenter applications (e.g., Web search) are sensitive to the tail of FCT and not mean, \name makes the right trade-off by 
prioritizing tail packets over average (or median) packets. Nevertheless, we expect to see better average FCT for PIAS because it emulates SJF, which is known to minimize average FCTs.

\subsubsection*{Throughput}
To see the performance of {\name} for long flows we measure {\em average} throughput in our simulations. Schemes that try to mimic SJF usually suffer from throughput issues; because they give strict priority to shorter flows. On the other hand, \name's mechanism to expedite ECN marked packets would speed up the congestion reaction processes at end host which increases the control over sender's rate. Figure \ref{fig:Through99} shows the average throughput of \name and PIAS for long flows. We see that {\name} achieves higher throughput for long flows in contrast to PIAS. Overall, {\name} achieves an average increase in long flows throughput
by about 32\%. 

\subsection{Queue length} 
Any packet scheduling scheme which keeps lower amount of packets in queues can successfully decrease the risk of packet drops in case of congestion. Recall from section \ref{sec:Sneaker} that since {\name} works in conjunction with DCTCP, it tends to store lower amount of packets in queues because of its nature which transmits congested packets faster. More specifically, By expediting congested tail packets, end host's transport protocol would receive congestion signals faster and consequently it cuts its Congestion Window (CWND) faster to speed up congestion recovery processes. 

\begin{figure}
\centering
\begin{subfigure}{.24\textwidth}
  \centering
  \includegraphics[width=\linewidth]{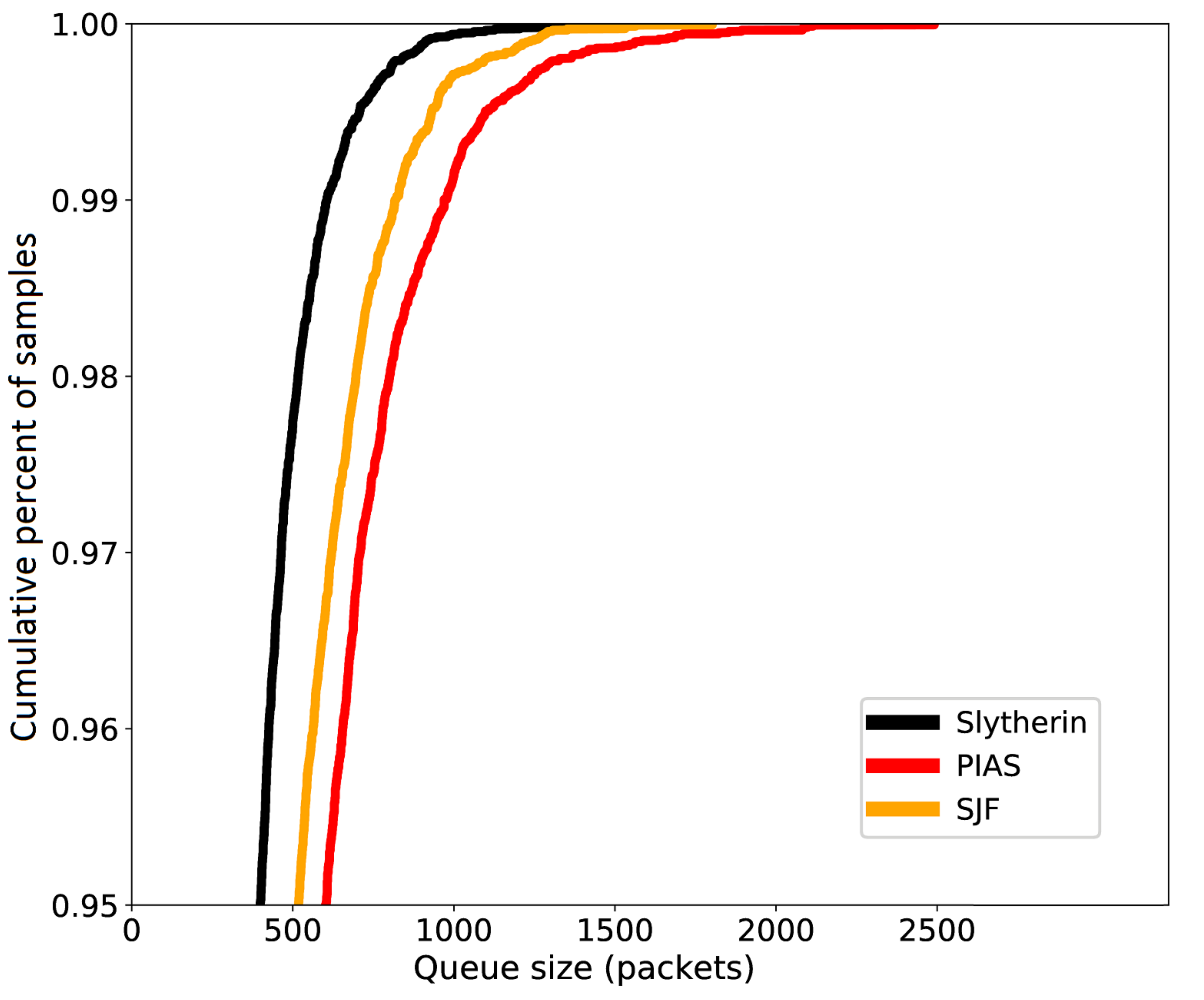}
  \caption{Load=40\%}\label{fig:queue1}
\end{subfigure}%
\begin{subfigure}{.24\textwidth}
  \centering
  \includegraphics[width=\linewidth]{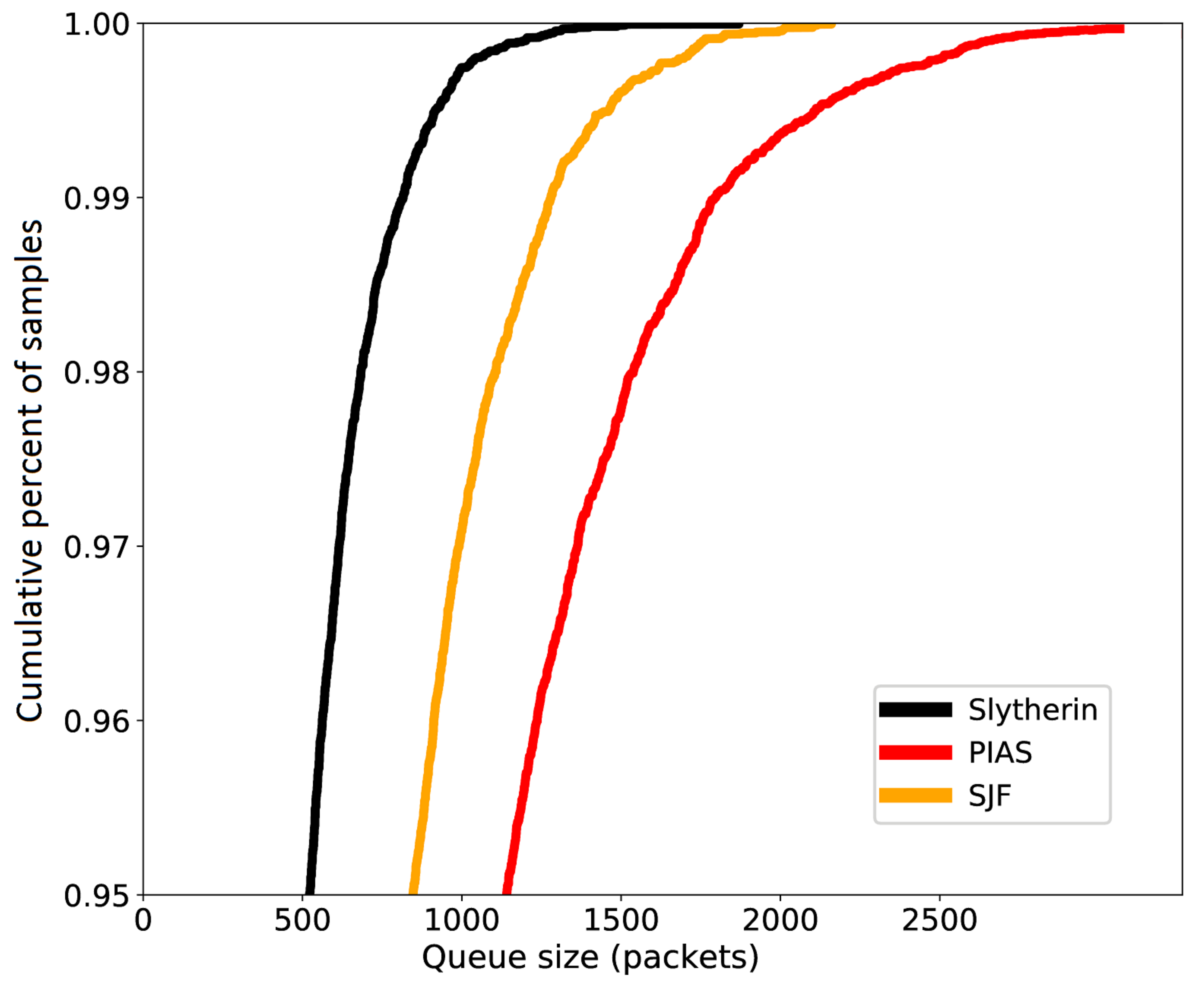}
  \caption{Load=60\%}\label{fig:queue2}
\end{subfigure}
\caption{CDF of queue length in switches}\label{fig:tailcdf}
\end{figure}

In figure \ref{fig:tailcdf}, we show the Cumulative Distribution Function (CDF) of queue lengths of all switches for \name, PIAS and Ideal SJF, for 40\% and 60\% load. 
In ideal SJF, we assume that we know flow sizes apriori and 
switches can, therefore, faithfully implement 
SJF. 

Figure \ref{fig:queue2} shows with a higher load factor (60\%), the stored number of packets in queues increases
for all schemes but \name outperforms others even more. 
Overall, \name significantly reduces $99^{th}$ percentile queue length in switches by about a factor of 2x on average.
While SJF is known to substantially improve \textit{average} flow completion times, our comparison with ideal SJF highlights \name's ability to 
\textit{specifically} target and improve tail latency beyond SJF. 

\begin{figure}[!htpb]
\centering
\includegraphics[width=0.5\linewidth]{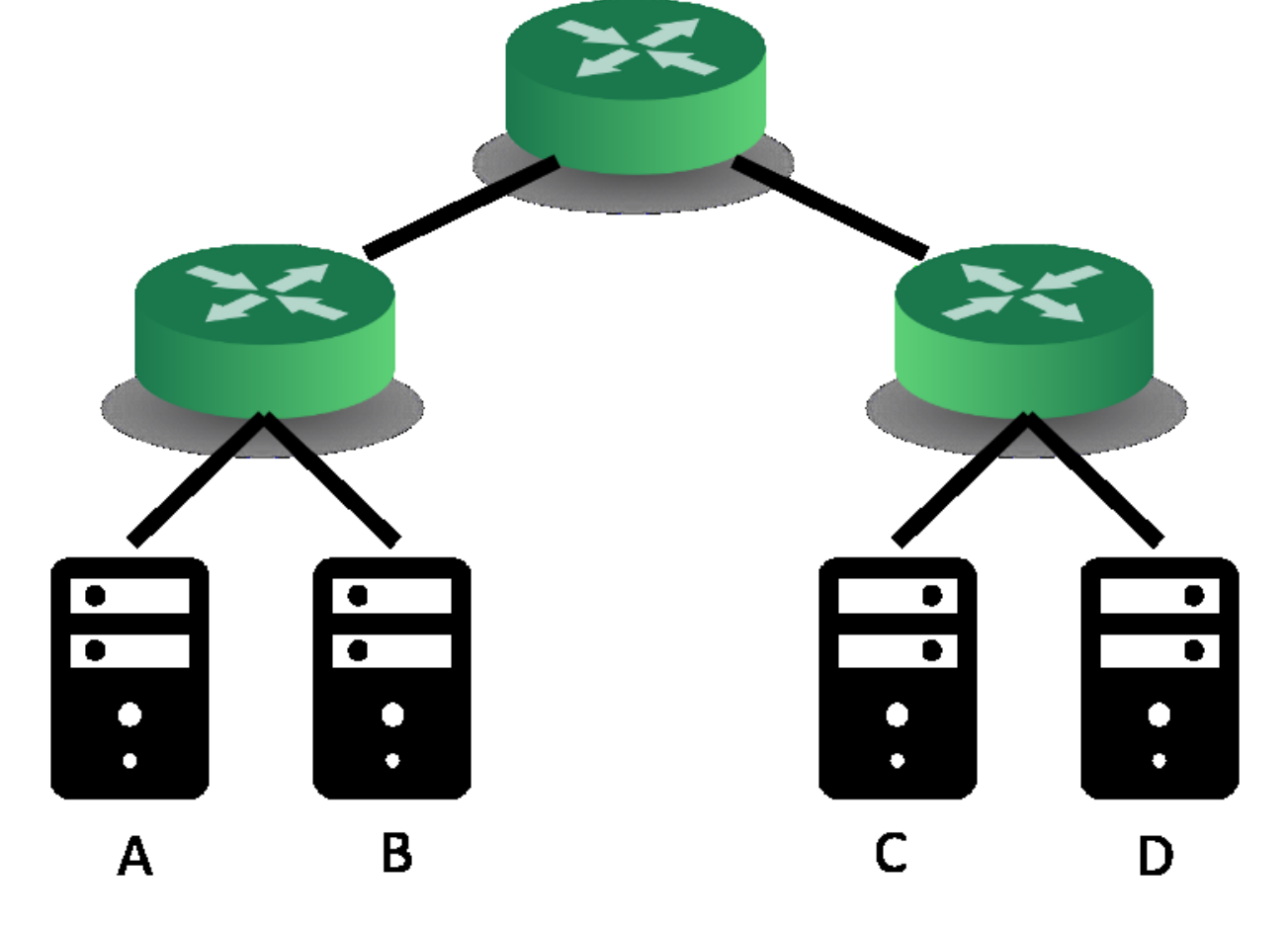}
\caption{Scenario used for convergence time evaluation}
\label{fig:topology}
\end{figure}

\subsection{Convergence time}
Expediting ECN marked packets 
helps TCP flows to converge faster to their fair share bandwidths 
when multiple flows compete on a bottleneck link. 
We evaluate \name's convergence using a small network (Figure~\ref{fig:topology})
consisting of four servers. 
We initiate a long flow from host A to host C and then, after 10 RTTs, 
we start 20 concurrent flows from server B to server D.

\begin{figure}
\centering
\includegraphics[scale=0.45,clip]{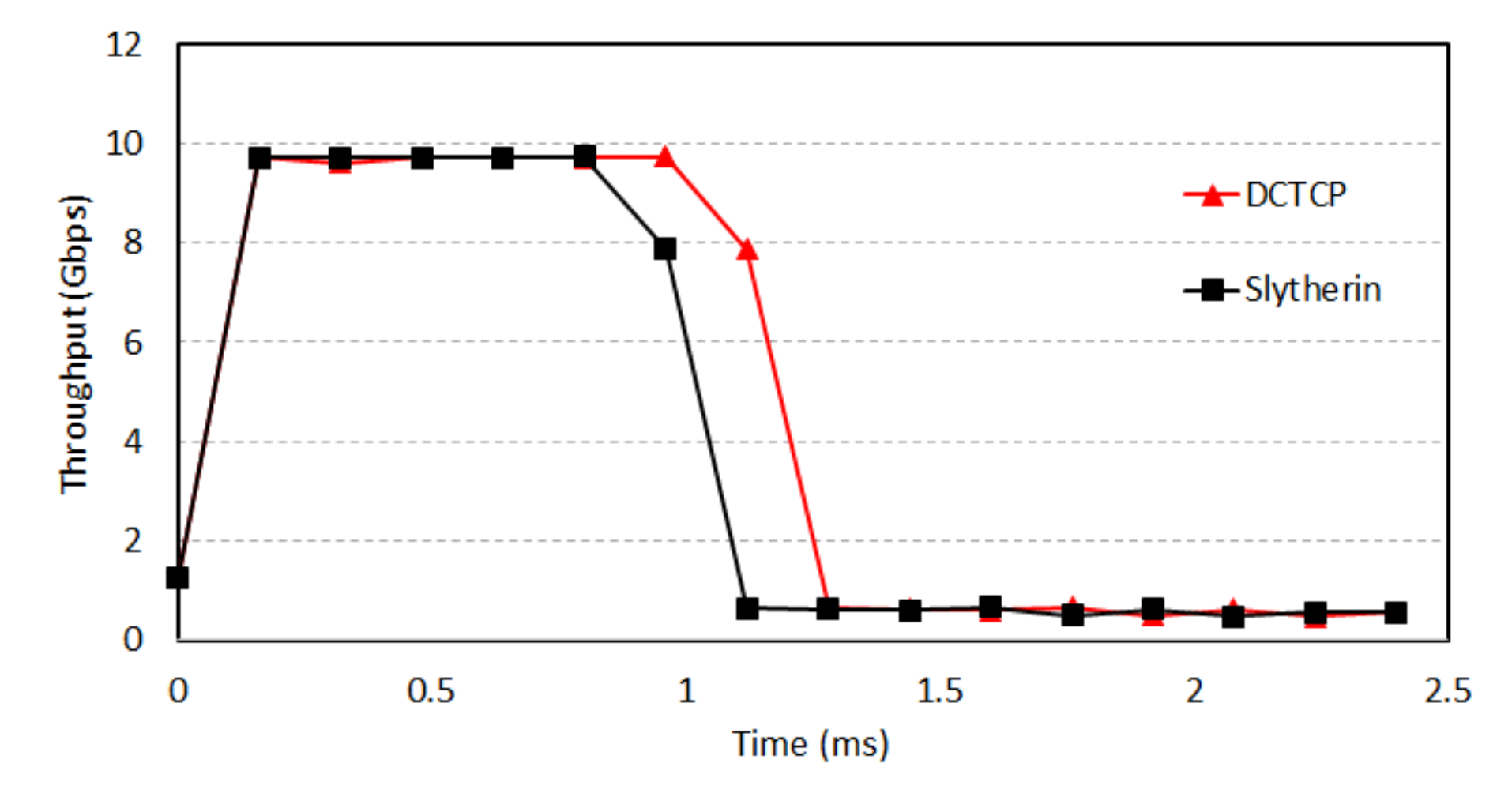}
\caption{Convergence time} \label{fig:convergence1}
\end{figure}

Figure \ref{fig:convergence1} shows the convergence time (i.e., time to 
reach the fair share rate) of DCTCP (baseline) and \name; X-axis 
shows time and Y-axis shows throughput (measured per each RTT).
We see that DCTCP takes 6 RTTs (480 ${\mu}s$) to converge to fair share 
whereas \name converges in 4 RTTs  (320 ${\mu}s$). 
Because \name provides faster convergence, 
it effectively mitigates utilization and fairness issues in multi-bottleneck 
scenarios, as reported in prior studies~\cite{cho2017credit}.

\subsection{Sensitivity to incast degree}
\begin{figure}
\centering
\includegraphics[scale=0.48,clip]{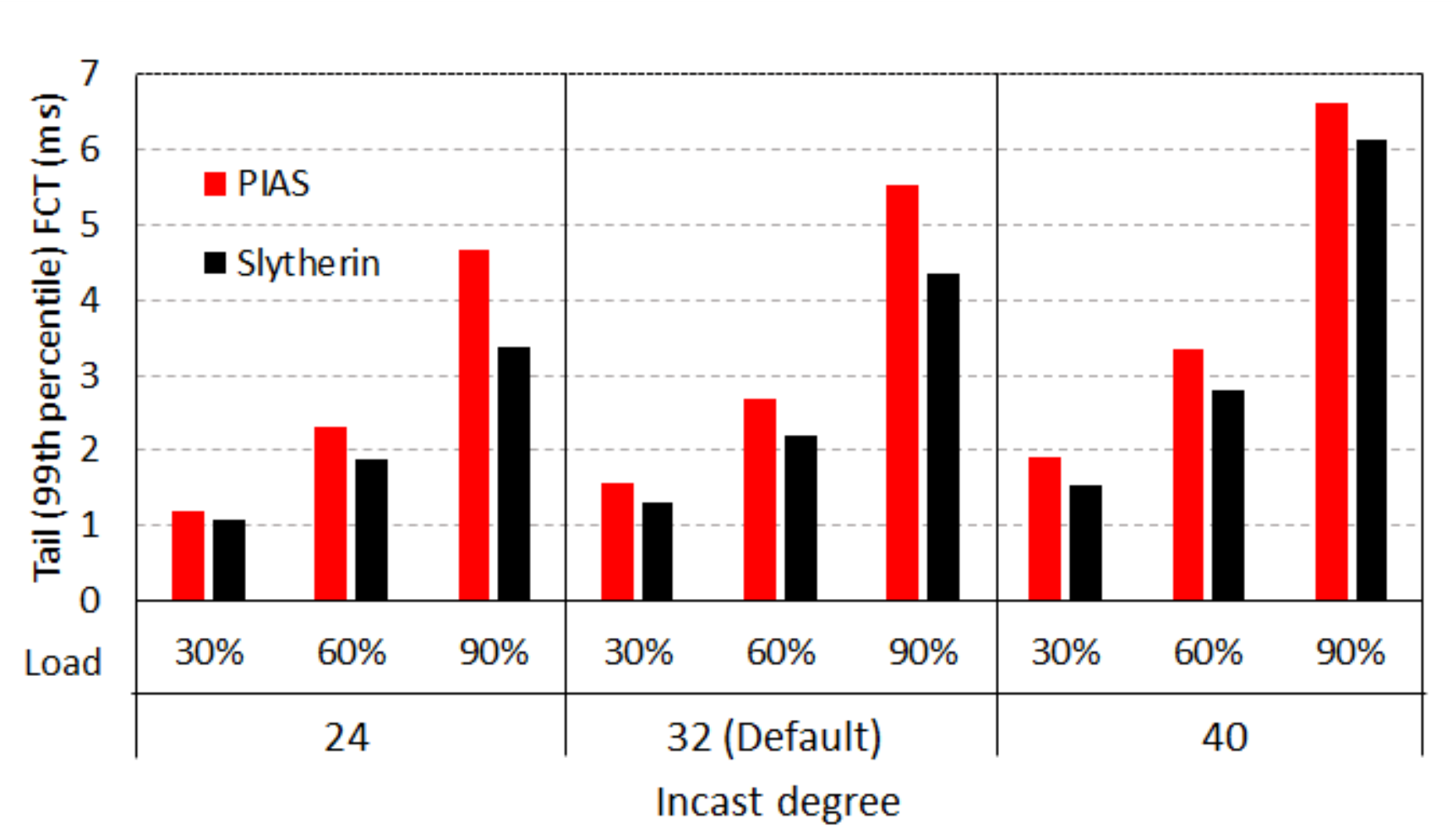}
\caption{Sensitivity to incast} \label{fig:incast}
\end{figure}

We study {\name}'s sensitivity to incast degree and compare its 
tail flow completion times to those of PIAS. 
Figure \ref{fig:incast} shows the $99^{th}$ percentile 
flow completion times of \name and PIAS as we vary the 
incast degree as 24, 32 (our default), and 40 along X-axis, for 
different loads. 
As expected, the flow completion times increase with increasing incast degree and load. 
\name relative gains are robust across incast degrees and loads. 
Overall, {\name} achieves an average reduction in $99^{th}$ percentile FCT
by about 21\%. 

\subsection{Sensitivity to ECN threshold}
\begin{figure}
\centering
\includegraphics[scale=0.45,clip]{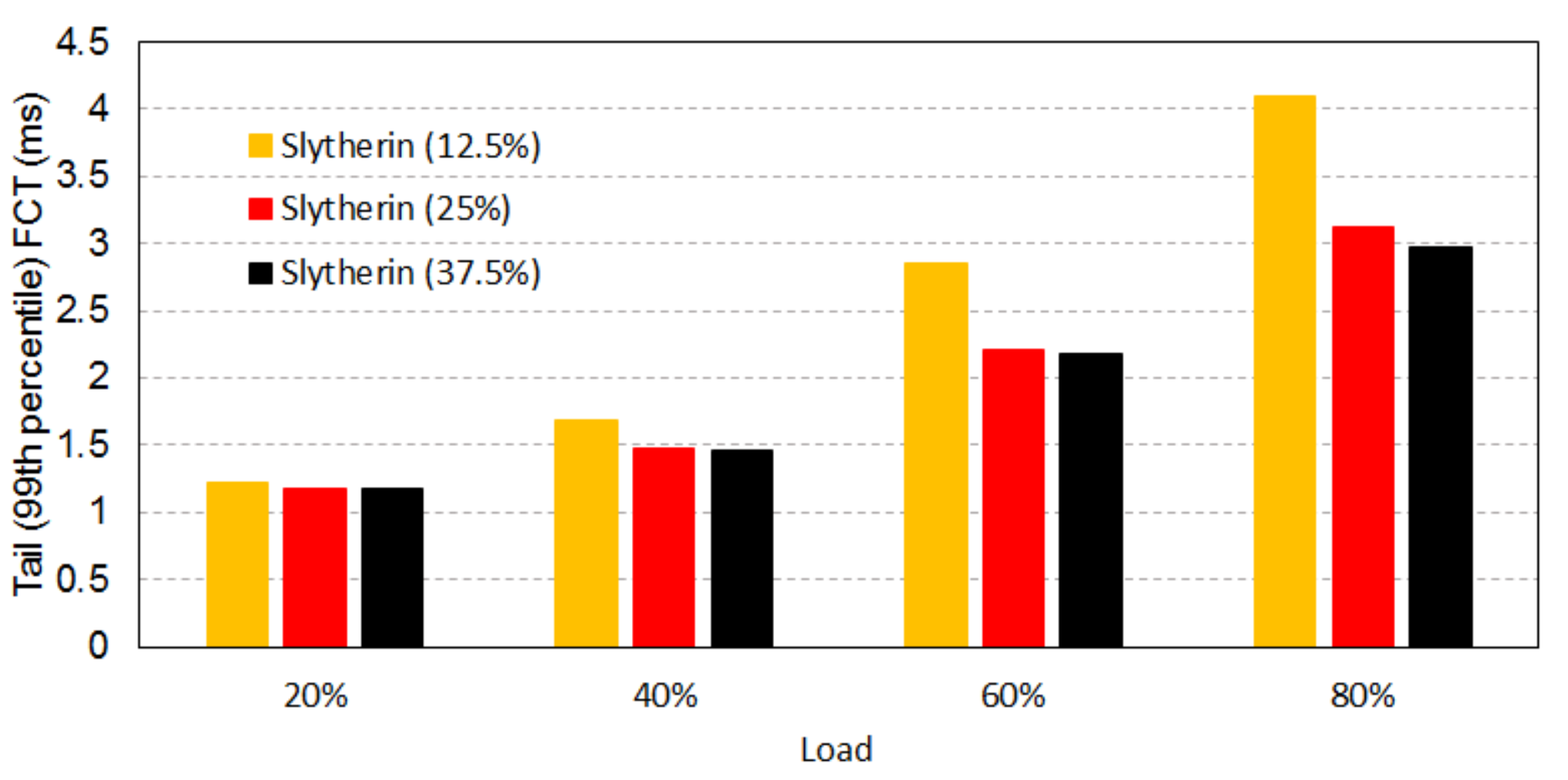}
\caption{Sensitivity to threshold} \label{fig:sens2thresh}
\vspace{-0.2in}
\end{figure}

Next, we analyze the \name's sensitivity to ECN threshold. 
While a lower threshold would aggressively mark packets, promote 
more packets to the high priority queue, and cause congestion, 
a higher threshold would be slow to react to congestion. 
Figure \ref{fig:sens2thresh} shows the tail flow completion times of \name for 
three different ECN threshold values of 12.5\%, 25\%, and 37.5\% of total buffer size
for varying loads. 
We see that \name with threshold of 12.5\% of queue szie suffers at higher loads as more 
packets get promoted to high priority queue and cause congestion. 
While a larger threshold of 37.5\% of total buffer size achieves better (lower) $99^{th}$
percentile FCT at higher loads, we observed loss of throughput 
for long flows (not shown) due to slower reaction to congestion. 

\subsection{Packet Reordering}
In this section we investigate the effect of expediting ECN marked packets on TCP packet reordering. \name schedules congested packets ahead of others to decrease tail flow completion time which may cause packet reordering at the end hosts. We compare the number of reordered packets in \name and PIAS to check \name's performance in terms of packet reordering efforts. 
PIAS authors assume the switch has enough rooms in the buffer so that each of those sub-queues (e.g., priority queues) can accommodate all incoming packets. Furthermore, PIAS sets a priority for the whole flow and consequently the number of reordered packets could be nearly zero. However, in a more realistic scenario, when the capacity of each of those sub-queues is limited, PIAS needs to either drop or demote packets to a lower priority queue which both lead to significant packet reordering.

\begin{figure}
\centering
\includegraphics[scale=0.45,clip]{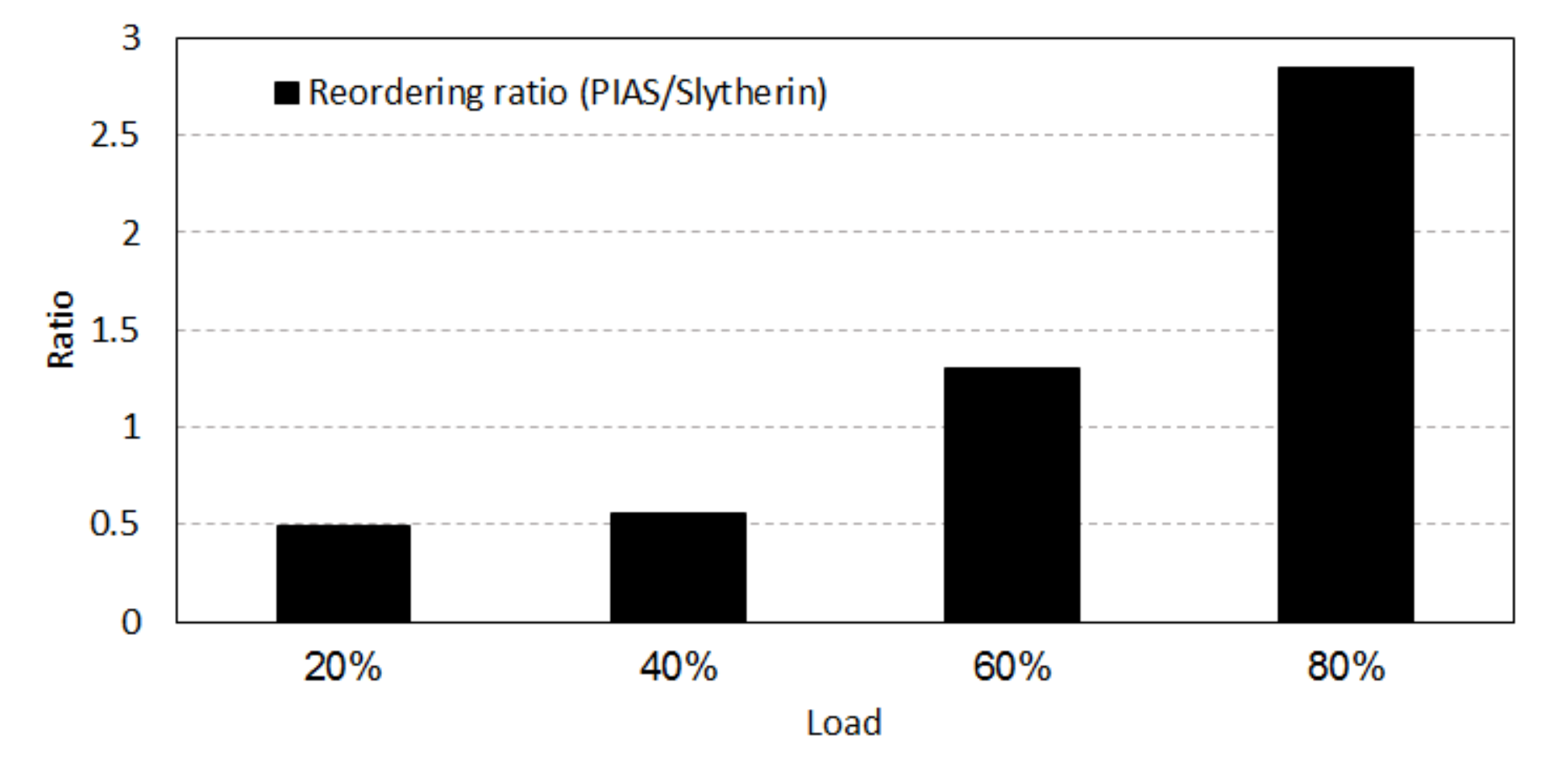}
\caption{Packet reordering ratio (PIAS/\name)}\label{fig:reordering}
\end{figure}

Figure \ref{fig:reordering} shows the number of reordered packets in \name compared to PIAS. X-axis shows the load factor on network and Y-axis shows the ratio of PIAS's number of reordered packets to {\name}. As shown in figure \ref{fig:reordering}, while PIAS reorders fewer packets than \name at low loads, PIAS incurs higher packet reordering 
than \name at high loads (i.e., at 60\% and higher loads). As load increases, PIAS demotes more packets. Our experiments show that while {\name} reorders only 0.56\% of total number of packets (on average), PIAS reorders 1.2\% of all packets (on average) across all loads. 
Because the absolute number of packet reordering would be far greater 
at high loads than at low loads, \name is more effective than PIAS in 
reducing reordering effort. 

\section{Related Work}
\label{sec:relatedwork}
There are many of past work that deal with the subjects of datacenter flow scheduling. These schemes usually require hardware modifications or rely on prior knowledge about the flows. A comprehensive review of all past proposals is beyond the scope of this paper, but we summarize some of the most relevant work here.

Earliest Deadline First (EDF) \cite{liu1973scheduling} is one of the oldest packet scheduling algorithms and is provably optimal when flow deadline is tagged on each single packet. D3 \cite{d3} suggests to assign different rates to flows based on their sizes and deadlines, but $D^{2}TCP$ \cite{d2tcp} and MCP \cite{mcp} both try to provide deadline-aware ECN-based congestion window adjustment.

Some other methods try to use prior information about flow sizes. pFabric \cite{pfabric} and PDQ \cite{pdq} both try to schedule packets based on flow size or flow remaining size so that the shortest flow will get the higher priority. On the other hand, other schemes like \cite{munir2013minimizing} try to use indirect methods to assign different priorities to flows. As an instance, HULL \cite{hull} tries to improve the speed of DCTCP's congestion window adjustment by trading bandwidth. While some methods like PASE \cite{munir2015friends} use Shortest Remaining Processing Time (SRPT), other schemes like UPS \cite{ups} aim to minimize FCT by leveraging Least Slack Time First (LSTF) techniques that seem to be nearly optimal.

There are many other schemes that can not be considered as packet scheduling algorithms but they still try to minimize FCT. For example, Rate Control Protocol (RCP) \cite{rcp} can achieve significant improvement in FCT of short flows. RCP is nearly optimal if minimizing FCT is the only metric for evaluation performance. RCP replaces TCP’s slow start mechanism with an alternative approach that allocates fair share bandwidth to all flows at the bottleneck links. Similar to D3, RCP requires hardware modifications at switches which makes it difficult to implement. QCN \cite{qcn} is a congestion control scheme that proposes a new method to improve performance in datacenters by sending congestion feedback from switches to end hosts. By utilizing intelligent switches and the new reaction logic in the end host NICs, QCN reduces recovery times during congestion; but it still suffers from implementation issues. VCP \cite{vcp} is another similar scheme that relies on a mechanism like ECN feedback.

\section{Conclusion}
\label{sec:conclusion}

We presented \name, which identifies tail packets
and prioritizes them in the network switches. 
Unlike prior approaches that emulate SJF scheduling of 
packets which is well-known to minimize average flow 
completion times, \name optimizes tail flow completion 
times, a metric that is more relevant for many online
datacenter applications (e.g., Web search, Facebook). 
\name does not require extensive support for 
identifying tail packets and leverages already 
available congestion signals (i.e., ECN). 
Using realistic workloads on typical datacenter 
network topologies, we showed that \name reduces tail 
flow completions by about 18\% as compared to existing 
state-of-the-art schemes. Further, we also showed that 
\name drastically cuts the queue lengths and speeds up 
convergence.
We plan to 
investigate  how to further improve 
\name's accuracy in identifying tail packets
and work on efficient 
switch hardware implementations in the future. 
As data continues to grow at a rapid rate, 
schemes such as \name that minimize network tail latency 
will become even more important.


%\section*{Acknowledgment}




\bibliographystyle{IEEEtran}
\bibliography{bare_conf}
%\begin{thebibliography}{1}


%\end{thebibliography}

%\bibliographystyle{myIEEEtran}
%\bibliography{bare_conf}

\end{document}